\documentclass[review]{elsarticle}

\usepackage{color}

\journal{Energy Conversion and Management}

\begin{document}

\begin{frontmatter}

\title{Small-signal model for frequency analysis of thermoelectric systems}

\author[Prevert]{Y. Apertet \corref{mycorrespondingauthor}}
\cortext[mycorrespondingauthor]{Corresponding author}
\ead[url]{yann.apertet@gmail.com}
\author[Skoltech]{H. Ouerdane}

\address[Prevert]{Lyc\'ee Jacques Pr\'evert, F-27500 Pont-Audemer, France}
\address[Skoltech]{Center for Energy Systems, Skolkovo Institute of Science and Technology, 3 Nobel Street, Skolkovo, Moscow Region 143026, Russia}

\begin{abstract}
We show how small-signal analysis, a standard method in electrical engineering, may be applied to thermoelectric device performance measurement by extending a dc model to the dynamical regime. We thus provide a physical ground to \textit{ad-hoc} models used to interpret impedance spectroscopy of thermoelectric elements from an electrical circuit equivalent for thermoelectric systems in the frequency domain. We particularly stress the importance of the finite thermal impedance of the thermal contacts between the thermoelectric system and the thermal reservoirs in the derivation of such models. The expression for the characteristic angular frequency of the thermoelectric system we obtain is a generalization of the expressions derived in previous studies. In particular, it allows to envisage impedance spectroscopy measurements beyond the restrictive case of adiabatic boundary conditions often difficult to achieve experimentally, and hence \emph{in-situ} characterization of thermoelectric generators. 
\end{abstract}
\end{frontmatter}


\section{Introduction}
Because of their small energy conversion efficiency thermoelectric solutions are typically less advantageous compared to other power-production technologies over a wide range of temperatures\cite{Vining2009}. However, as thermoelectric systems do not require moving parts to operate, their reliability is higher than those of standard heat engines and, as thermoelectric energy conversion is essentially electronic in nature, thermoelectricity provides a convenient way to directly convert waste heat into electric power. So, in addition to the great efforts made for the improvement of device and materials performance, focusing especially on nanoscale structures \cite{Heremans2013,Mori2013}, much work is also devoted to finding innovative solutions for the integration of thermoelectric devices in combined generation and refrigeration cycles\cite{Shu2012,Yilbas2014,Zheng2014,Zhang2015}. For these latter schemes, it is obviously crucial to obtain reliable measurements of the overall efficiency to assess their technological and economic interest \cite{Yee3013,LeBlanc2014}, but one of the challenges to be met is to properly and accurately determine the performance of the thermoelectric device itself, which can be related to the theoretical maximum efficiency of a thermoelectric generator working between two thermal reservoirs at temperatures $T_{\rm h}$ and $T_{\rm c}$ respectively, with $T_{\rm c}<T_{\rm h}$ \cite{Ioffe,Champier2017}: 

\begin{equation}\label{etamax}
\eta_{\rm max} = \eta_{\rm C} \frac{\sqrt{1+ZT} - 1 }{\sqrt{1+ZT} + T_{\rm c}/T_{\rm h}} 
\end{equation}

\noindent with $\eta_{\rm C} = 1-T_{\rm c}/T_{\rm h}$ being the Carnot efficiency, and $ZT$ the figure of merit being given by

\begin{equation}\label{ZT}
ZT = \frac{\alpha^2 \overline{T}}{R K_0} 
\end{equation}

\noindent where $\overline{T}$ is the average working temperature of the system, $R$ is the electrical resistance of the module, $K_0$ is its thermal conductance at vanishing electrical current, and $\alpha$ is the global Seebeck coefficient characterizing the thermoelectric coupling between the electrical current and the heat flux across the legs of the device. 

So thermoelectric performance boils down to the determination of a single quantity: $ZT$, but its accurate evaluation is far from being straightforward, and various approaches may apply: One may measure $\alpha$, $R$, and $K_0$, separately and then compute $ZT$ using Eq.~(\ref{ZT}). This method however proves quite inaccurate without great experimental care as each measurement error for each parameter contributes to the cumulated global error on the resulting value of $ZT$ \cite{Tritt1997,Zhou2005,Wang2015,Knud2016}. To overcome the unavoidable drawbacks of multiple measurements, Harman suggested that $ZT$ might be determined by only measuring, under adiabatic conditions, the voltage across the sample resulting from alternating current; this technique is known as the transient Harman method \cite{Harman1958}. Lisker later extended the idea of a single parameter measurement for different conditions: He proposed to measure electrical conductivity under adiabatic and isothermal conditions or, equivalently, to measure thermal conductivity under vanishing electrical current and short-circuit conditions \cite{Lisker1966}. Min and Row suggested that $ZT$ might also be evaluated by measuring the temperature difference across the sample for short-circuit and open-circuit conditions if the incoming heat flows remain constant during the experiment \cite{Min2001}. 

Recently, the Harman technique has been further improved using impedance spectroscopy analysis to lower uncertainties on $ZT$ \cite{DeMarchi2011}. With this technique, the frequency dependence of the voltage-current ratio across the system is analyzed and then compared to theoretical models. Besides the determination of $ZT$, it is also possible to gain additional information on the material's thermoelectric properties \cite{Downey2007,Hatzikraniotis2010,Garcia2014,Garcia2014JEM,Garcia2016}. To evaluate the different characteristics of the sample, measurements are fitted to an equivalent electrical model. Amongst the different proposed equivalent models, the simplest one corresponds to an RC electrical circuit \cite{Downey2007}. This latter was obtained empirically from measurements. In this article, we show how to associate this ac model with existing dc models for thermoelectric generators in order to interpret impedance spectroscopy measurements of thermoelectric modules explicitly accounting for their non-ideal coupling to heat source and sink. Indeed the presence of the finite thermal conductance of heat exchangers influences greatly the performance of thermoelectric devices, as shown, e.g., in optimization and performance improvement studies involving the thermal coupling to the heat sink in particular \cite{Elghool2017,Araiz2017}. The article is organized as follows. In Sec.~\ref{model}, we first present the dc model for thermoelectric generators including non-ideal thermal contacts. We then extend this model to the frequency domain using small-signal modeling. In Sec.~\ref{discussion}, we apply numerically our small-signal analysis to a commercial device and we discuss our approach and compare our results to the literature stressing similarities and discrepancies with other models.

\section{\label{model}Model}

\subsection{Classical model with non-ideal thermal contacts}

\begin{figure}
	\centering
		\includegraphics[width=0.75\textwidth]{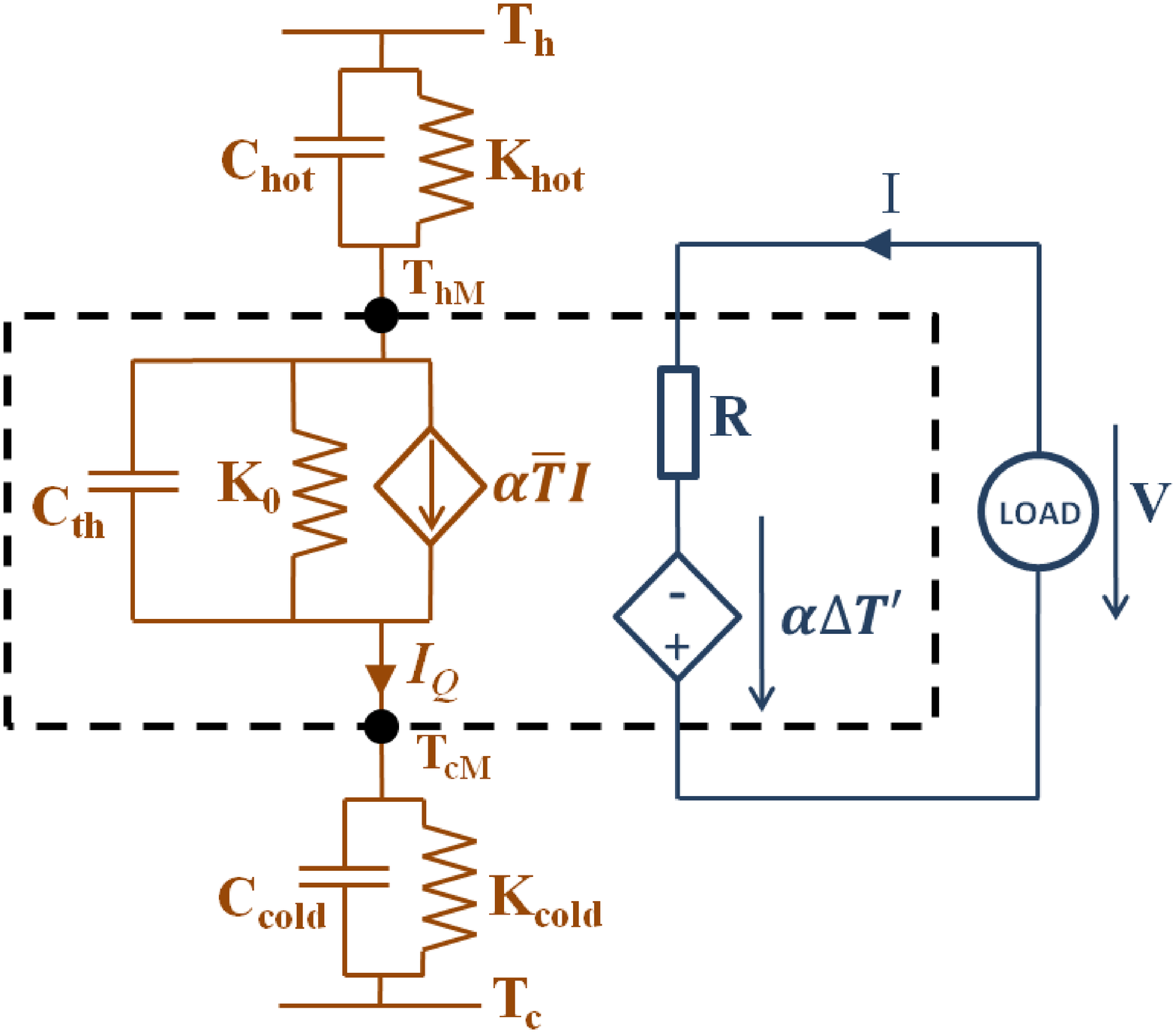}
	\caption{Schematic representation of the thermoelectric converter with non ideal thermal contacts.}
	\label{fig:figure1}
\end{figure}

We consider a thermoelectric generator connected through non-ideal thermal contacts to two thermal reservoirs at constant temperatures $T_{\rm h}$ and $T_{\rm c}$ respectively. The wording ``thermal contacts'' implies all the parts of the actual system that conduct the heat flux including the heat exchangers, heat conductive paste, ceramic layers, and copper stripes. With no loss of generality, we then assume that the thermal contacts thus defined are characterized by a finite thermal conductance $K_{\rm hot}$ (resp. $K_{\rm cold}$) on the hot (resp. cold) side. The electrical and thermal properties of this system in the dc regime are given by \cite{Apertet2012}:

\begin{eqnarray}
V = \alpha \Delta T' - RI, \label{Thevenin}\\
I_{Q} = \alpha \overline{T} I + K_0 \Delta T'\label{heatcurrent}
\end{eqnarray}

\noindent where $V$ and $\Delta T'$ are respectively the voltage and the temperature difference across the generator and $I$ and $I_Q$ are respectively the electrical current and the thermal current flowing through the device. To obtain the above equations, the effect of Joule heating is neglected and the thermal current is assumed to be constant along the device even if, actually, it slightly varies due to the energy conversion process. This assumption is quite reasonable in the generator regime under working conditions often met in practice when the temperature difference across the device, $\Delta T'$, is usually much smaller than the average temperature across the device $\overline{T}$, which is the case during, e.g., characterization at ambient temperature. 
It also greatly simplifies the model as $I_Q$ is then only composed of two distinct contributions: A convective heat current $\alpha \overline{T}I$ associated with the global displacement of the electrons \cite{Thomson1856,Apertet2012JPCS} and a conductive heat current $K_0 \Delta T'$ associated with heat leaks. Note that the term \emph{electric convection of heat}, in opposition to traditional heat conduction, has been coined in 1856 by Kelvin in Ref.\cite{Thomson1856}. Furthermore, we assume that the three thermoelectric parameters $R$, $K_0$ and $\alpha$ are constant and that the system exchanges heat only with the thermal reservoirs. The global system is described on Fig.~\ref{fig:figure1}. For completeness, the thermal capacitances of the thermoelectric generator, $C_{\rm th}$, and of the thermal contacts, $C_{\rm hot}$ and $C_{\rm cold}$, are also displayed even if they have no influence in the dc regime as they are then equivalent to a thermal open-circuit. 

Even if it might be suggested by its form, Eq.~(\ref{Thevenin}) does not correspond to a genuine Th\'evenin modeling of the thermoelectric generator since the electromotive force $\alpha \Delta T'$ depends on the electrical current $I$ \cite{Apertet2012}. This point is stressed by the use of a controlled voltage source rather than an ideal voltage source in Fig.~\ref{fig:figure1}. This dependence is due to the combined effects of the convective heat current and of the non-ideal thermal contact conductances: Because of the thermal contacts, the temperature difference $\Delta T'$ across the generator is different from the temperature difference between the thermal reservoirs $\Delta T$, the latter being constant. Since we assume constant heat current $I_Q$ across the device, the thermal circuit might be seen as a thermal equivalent of a voltage divider and one then gets: 

\begin{equation}\label{DeltaTprime}
\Delta T' = T_{\rm hM} - T_{\rm cM} = \Delta T - \frac{I_Q}{K_{\rm C}} 
\end{equation}

\noindent where $K_{\rm C}$ is the equivalent thermal conductance of the contacts, given by $K_{\rm C} = K_{\rm hot} K_{\rm cold}/(K_{\rm hot}+K_{\rm cold})$. Due to the conductive heat current, a change on the electrical current $I$ thus has consequences on the temperature difference $\Delta T'$ and hence on the electromotive force $\alpha \Delta T'$. It is then possible to obtain a genuine Th\'evenin model of this thermoelectric generator distinguishing constant terms from terms depending on the electrical current $I$. The voltage output thus reads 

\begin{equation}\label{RealThevenin}
V = \frac{K_{\rm C}}{K_0 + K_{\rm C}}\alpha \Delta T  - \left(R + \frac{\alpha^2 \overline{T}}{K_0 + K_{\rm C}}\right) I
\end{equation}
\noindent where the first term of the right hand side is the actual electromotive force \cite{Apertet2012}. The additional resistive term stems from the introduction of the non-ideal thermal contacts.


\subsection{Small-signal electrical model}

To perform a potentiostatic impedance spectroscopy measurement, a small ac voltage with sweeping frequency is applied using a lock-in to a circuit made of a sense resistor and of the thermoelectric generator (see, e.g., Ref.~\cite{Downey2007}); the resulting ac current in the circuit $\delta I$ and the voltage $\delta V$ are both measured and then used to compute the complex impedance ${\mathcal Z}$ of the circuit:
\begin{equation}
{\mathcal Z} = - \frac{\delta V}{\delta I}.
\end{equation}

\noindent Note that galvanostatic impedance spectroscopy, performed by application of a small ac current in the circuit and measurement of the resulting voltage across the system, can also be used to obtain the impedance.

We make the standard assumption that each variable $X$ in the circuit may be defined as the sum of a continuous contribution and of a varying contribution at a frequency imposed by the lock-in: 

\begin{equation}\label{decompo}
X = X_{\rm dc} + \delta X
\end{equation}

\noindent with the value of the dc component being set by the choice of the working conditions. Hence, the impedance ${\mathcal Z}$ is related \emph{only} to the ac part of each variable, typically small compared to their dc counterpart. Since we focus only on these ac contributions, the associated model in which dc biases are ignored, is termed \emph{small-signal model}. The dc model may then be modified to keep only these contributions. To do so, and hence to obtain the small-signal electrical model of the circuit, one needs to shut down every non controlled dc source \cite{Boylestad2013}. In the model displayed on Fig.~\ref{fig:figure1}, the only non controlled dc source is encountered in the thermal part of the device. Indeed, the constant temperature difference between the thermal reservoirs might be seen as a thermal potential source (by analogy with a voltage source) and consequently has to be replaced by a thermal short circuit between the thermal reservoirs; This amounts to discarding the dc contribution to the temperature difference across the system. Since in this case both thermal contact impedances are in series, one gets the schematic representation of the small signal model given in Fig.~\ref{fig:figure2}. One may then notice that the complex impedance ${\mathcal Z}$ may be split into two distinct parts: 

\begin{figure}
	\centering
		\includegraphics[width=0.75\textwidth]{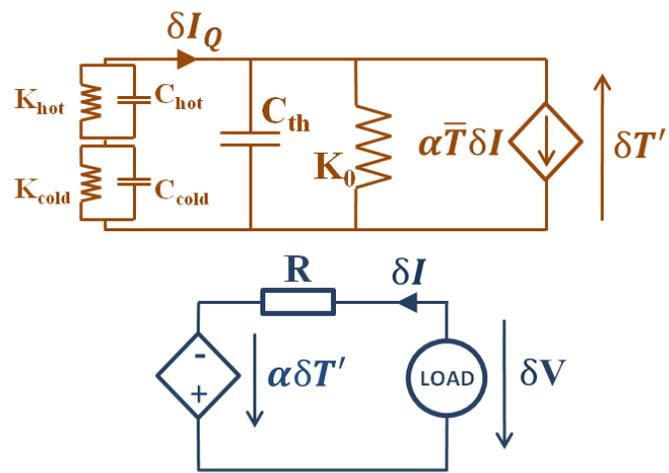}
	\caption{Small-signal model of the thermoelectric converter with non-ideal thermal contacts. The upper part is the thermal circuit while the lower part is the electrical circuit.}
	\label{fig:figure2}
\end{figure}

\begin{equation}\label{Z}
{\mathcal Z} = R + {\mathcal Z}_{\rm th} 
\end{equation}

\noindent where ${\mathcal Z}_{\rm th}$ is the electrical impedance stemming from the thermal part of the circuit, defined as: 

\begin{equation}
{\mathcal Z}_{\rm th} = - \frac{\alpha \delta T'}{\delta I}.
\end{equation} 

The small temperature variation due to the varying electrical load is obtained from the thermal part of Fig.~\ref{fig:figure2}. Indeed, we see that the thermal admittance is given by: 

\begin{eqnarray}
- \frac{\alpha \overline{T}\delta I}{\delta T'} & = & jC_{\rm th}\omega + K_0\\
\nonumber 
&+& \frac{\left(K_{\rm hot} + jC_{\rm hot}\omega\right)\left(K_{\rm cold} + jC_{\rm cold}\omega\right)}{K_{\rm hot} + jC_{\rm hot}\omega + K_{\rm cold} + jC_{\rm cold}\omega} 
\end{eqnarray}

\noindent with $j^2 = -1$. In order to keep compact notations, we define the global thermal capacitance of the contacts as: 

\begin{equation}
C_{{\rm C}} = \frac{C_{\rm hot}C_{\rm cold}}{C_{{\rm hot}}+C_{{\rm cold}}}. \nonumber
\end{equation} 

\noindent Moreover, in order to reflect the asymmetry between the thermal contacts on the hot side and on the cold side, we introduce two contrast functions: 

\begin{equation}
\Psi_{\rm K} = \frac{K_{{\rm hot}} - K_{{\rm cold}}}{K_{{\rm hot}}+K_{{\rm cold}}}\nonumber
\end{equation}

\noindent and 

\begin{equation}
\Psi_{\rm C} = \frac{C_{{\rm hot}} - C_{{\rm cold}}}{C_{{\rm hot}}+C_{{\rm cold}}}\nonumber
\end{equation}

\noindent The former reflects the asymmetry on the thermal conductances while the latter reflects the asymmetry on the thermal capacitances. These two quantities thus vary between $-1$ and $1$ and vanish for symmetric contacts. 
Tedious but simple calculations lead to the following canonical expression for the complex impedance ${\mathcal Z}_{\rm th}$: 
\begin{equation}\label{Zth}
{\mathcal Z}_{\rm th} = R_{\rm TE}\frac{1+j\omega/\omega_{\rm 1}}{1 + 2 \zeta j\omega/\omega_{\rm 0} + \left(j\omega/\omega_{\rm 0}\right)^2},
\end{equation} 

\noindent where the dc value $R_{\rm TE}$ of this complex impedance is given by  

\begin{equation}\label{RTE}
R_{\rm TE} =  \frac{\alpha^2 \overline{T}}{K_0 + K_{\rm C}} 
\end{equation} 

\noindent the characteristic angular frequencies are 

\begin{equation}
\omega_{\rm 0} = \sqrt{\frac{K_{\rm C}}{C_{\rm C}}\left(\frac{K_{{\rm 0}}+K_{{\rm C}}}{C_{{\rm th}}+C_{{\rm C}}}\right)\left(\frac{1-{\Psi_{\rm C}}^2}{1-{\Psi_{\rm K}}^2}\right)} \nonumber
\end{equation} 

\noindent and 

\begin{equation}
\omega_{\rm 1} =\frac{K_{\rm C}}{C_{\rm C}}\left(\frac{1-{\Psi_{\rm C}}^2}{1-{\Psi_{\rm K}}^2}\right) \nonumber 
\end{equation}

\noindent and where the damping factor $\zeta$ is given by:

\begin{equation}
\zeta = \frac{1}{2\omega_{\rm 0}}\left[\frac{{\omega_{\rm 0}}^2+{\omega_{\rm 1}}^2}{\omega_{\rm 1}}+\frac{K_{\rm C}({\Psi_{\rm K}}-{\Psi_{\rm C}})^2}{(C_{{\rm th}}+C_{{\rm C}})(1-{\Psi_{\rm K}}^2)}\right].  \nonumber
\end{equation}

\noindent The total electrical impedance of the system might then be easily obtained combining Eqs.~(\ref{Z}) and (\ref{Zth}).


%

Interestingly, for symmetric contacts, i.e., ${\Psi_{\rm K} =\Psi_{\rm C}=0}$, Eq.~(\ref{Zth}) is greatly simplified and the total electrical impedance ${\mathcal Z}$ becomes 

\begin{equation}\label{Zsimple}
{\mathcal Z}_{\rm sym} = R + \frac{R_{\rm TE}}{1 + j\omega/\omega_{\rm TE}} 
\end{equation} 

\noindent where the characteristic angular frequency is 

\begin{equation}\label{omega}
\omega_{\rm TE} = \frac{{\omega_{\rm 0}}^2}{\omega_{\rm 1}} =  \frac{K_0 + K_{\rm C}}{C_{\rm th}+C_{\rm C}}.
\end{equation} 

This simplification is particularly valuable since this situation is often encountered in practice (e.g., both ceramic plates surrounding commercial thermoelectric modules are identical). Actually, Eq.~(\ref{Zsimple}) has even a broader validity than only symmetric situation since it also holds for the more general condition ${\Psi_{\rm K} =\Psi_{\rm C}}$, i.e., when the complex thermal impedance of the contact on the hot side is proportional to the complex thermal impedance of the contact on the cold side. It is possible to associate  Eq.~(\ref{Zsimple}) with a pure electrical circuit. In this case, the form of the characteristic frequency $\omega_{\rm TE}$ leads to the introduction of a so-called thermoelectric capacitance \cite{Garcia2014JEM}: 

\begin{equation}\label{CTE}
C_{\rm TE} =  \frac{1}{R_{\rm TE}\omega_{\rm TE}} = \frac{C_{\rm th}+C_{\rm C}}{\alpha^2 \overline{T}}.
\end{equation} 

\begin{figure}
	\centering
		\includegraphics[width=0.75\textwidth]{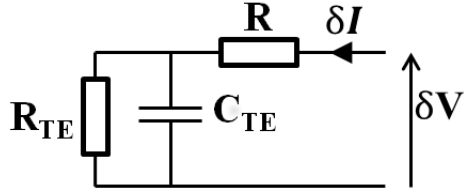}
	\caption{Equivalent electrical circuit corresponding to Eq.~(\ref{Zsimple}).}
	\label{fig:figure3}
\end{figure}

\noindent The equivalent electrical circuit is then a simple RC circuit given in Fig.~\ref{fig:figure3}.

\subsection{Illustrative example}

\begin{figure}
	\centering
		\includegraphics[width=0.75\textwidth]{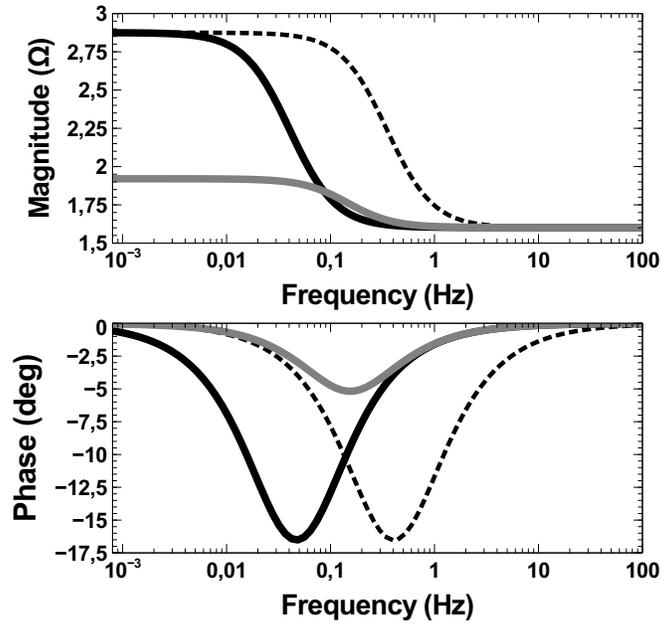}
	\caption{Bode diagrams of the impedance ${\mathcal Z}$ for a commercial thermoelectric module under different assumptions: Thermally insulated module (full black lines), module connected to thermal reservoirs through realistic thermal contacts (full gray lines) and thermally insulated module with neglected contact thermal capacitances (dotted lines).}
	\label{fig:figure4}
\end{figure}

To illustrate the impact of thermal contacts on the small-signal response of a thermoelectric module, we consider the commercial one (Kryotherm, TB-127-1.4-1.2) used by Lineykin and Ben-Yaakov in Ref.~\cite{Lineykin2005}. Its characteristic parameters are given in Table~\ref{table1}, but these alone are not sufficient to predict the module's response in frequency as this latter also depends on thermal contacts. Hence, we envisage two different situations. First, we assume a thermally insulated module, i.e., $K_{\rm C} = 0$; the thermal capacitance of the contacts is however taken into account. For simplicity, we consider that this thermal capacitance stems only from the ceramic plates and consequently that $C_{\rm C} =  2.67~$J.K$^{-1}$ \cite{Lineykin2005}. The expected dependence of the electrical impedance ${\mathcal Z}$ of this system on the frequency $f = \omega/(2\pi)$ is shown as a Bode diagram on Fig.~\ref{fig:figure4} (full black lines). The associated characteristic frequency $f_{\rm TE} = \omega_{\rm TE}/(2\pi)$ obtained from Eq.~(\ref{omega}) is $35.2~$mHz.

\begin{table}
	\centering
	\begin{tabular}{ccccc}
  \hline
	\hline
	~&~&~&~&~\\
  $R$ ($\Omega$)~~& $\alpha$ (V.K$^{-1}$)~~& $K_0$ (W.K$^{-1}$)~~& $C_{\rm th}$ (J.K$^{-1}$)~~& $ZT$ \\
	~&~&~&~&~\\
  1.602 & 0.0532 & 0.667 & 0.35 & 0.795\\
	~&~&~&~&~\\
  \hline
	\hline
\end{tabular}
	\caption{Characteristic parameters of a commercial thermoelectric module TB-127-1.4-1.2 (Kryotherm) at $\overline{T} = 300~K$ (from Ref.~\cite{Lineykin2005}).}
	\label{table1}
\end{table}

\begin{figure}
	\centering
			\includegraphics[width=0.75\textwidth]{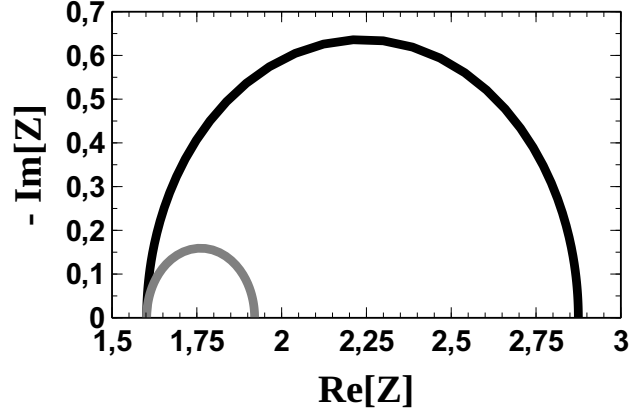}
	\caption{Nyquist plots of the impedance ${\mathcal Z}$ for a commercial thermoelectric module under different assumptions: Thermally insulated module (black line) and module connected to thermal reservoirs through finite-resistance thermal contacts (gray lines).}
	\label{fig:figure5}
\end{figure}

Next, we consider the same system except that it is no longer thermally insulated but connected to thermal reservoirs at ambient temperature ($\overline{T} = 300~$K), through contacts. The associated thermal conductance is set as $K_{\rm C} = 2~$W.K$^{-1}$. The electrical impedance behavior expected in this case is also shown on Fig.~\ref{fig:figure4} (full gray lines) and the characteristic frequency $f_{\rm TE}$ is then $0.141$~Hz. The comparison between these two cases is quite illustrative as the connexion with thermal reservoirs, which is always met when the module is actually used for energy conversion or for cooling applications, greatly modifies the response of the thermoelectric system. Indeed, two effects of the presence of finite thermal conductance contacts, $K_{\rm C}$, are to increase the characteristic frequency and to decrease the magnitude variation between the low- and high-frequency regions of the Bode diagram. Since this variation corresponds to $R_{\rm TE}$, this latter modification can easily be understood with Eq.~(\ref{RTE}). Considering the thermal contacts when interpreting impedance spectroscopy measurements thus appears as mandatory for generalizing this technique to in-situ measurements. The Nyquist plots, i.e., the imaginary part of the complex impedance ${\mathcal Z}$ as a function of the real part of ${\mathcal Z}$, for the three previous examples are displayed on Fig.~\ref{fig:figure5}. As expected, one recovers a half circle associated with Eq.~(\ref{Zsimple}) whose radius is linked to both thermal conductances of the module and of the contacts but not to the thermal capacitances. Hence, contrary to Bode diagrams, the Nyquist plot for a thermally insulated module is not modified when neglecting contact thermal capacitances.

\section{\label{discussion}Discussion}

\subsection{Physical origin of $R_{\rm TE}$ and $C_{\rm TE}$}
It is striking that the thermoelectric resistance $R_{\rm TE}$ assumes a form similar to the additional term for the electrical resistance appearing in Eq.~(\ref{RealThevenin}), i.e. the Th\'evenin expression for the dc voltage $V$, due to the finite thermal contact conductances. For both ac and dc regimes, this resistance stems from the modification of the electromotive force $\alpha \Delta T'$ with the electrical current. Indeed, with an increase of the electrical current, the thermal flux also increases because of the convective part of this latter. This leads to larger temperature difference between the thermal contacts. The temperature difference between the reservoirs $\Delta T$ being constant, the effective temperature difference applied across the thermoelectric legs is thus smaller. Consequently, a higher electrical current results in a lower effective electromotive force $\alpha\Delta T'$: The electrical resistance $R_{\rm TE}$ reflects this detrimental effect.

When the frequency increases sufficiently, the additional thermal contribution ${\mathcal Z}_{\rm th}$ to the total impedance vanishes. Indeed, in this case, the thermal contacts have no longer any influence: At high frequency, the thermal capacitances create a bypass for the heat current and impose $\delta T' =0$. Since this phenomenon is of pure thermal nature, the associated characteristic frequency $\omega_{\rm TE}$ only involves thermal properties [see Eq.~(\ref{omega})]. This thermal process leads to the appearance of the so-called thermoelectric capacitance $C_{\rm TE}$ in the equivalent electrical circuit. The typical value of this capacitance is about 1~F \cite{Downey2007,Garcia2014JEM}, which is several orders of magnitude higher than standard values for genuine electrical capacitances. The appearance of such capacitance and its surprisingly high value might however be explained referring to the thermoelectric convection: According to Thomson, the Seebeck coefficient might be related to the specific heat of each electrical carrier \cite{Thomson1854}. Since theses carriers transport at the same time heat and electrical charges, these two quantities are tightly coupled, which is as a matter of fact the footprint of thermoelectricity. It thus implies that evacuating electrical charges from a thermoelectric system is tightly linked to evacuating heat from this same system. Hence, the thermal capacity of the system might limit the convective heat flow and so the electrical current: This behavior is typical of an electrical capacitance. Moreover, since a thermal system has much higher time constant than an electrical system, the effective electrical capacitance constituted that way is larger than genuine electrical capacitances.

\subsection{Comparison with previous models}

There are several models to assess the frequency response of thermoelectric systems. The most basic model corresponds to an RC circuit obtained empirically \cite{Downey2007}. In this model equivalent to the one displayed in Fig.~\ref{fig:figure3}, Downey and coworkers also obtain a thermal contribution to the electrical impedance but they do not distinguish the contribution of thermal contacts from the contribution of the thermoelectric material. Despite its simplicity, this model is sufficient to describe the frequency response of commercial thermoelectric modules \cite{Downey2007}. However the RC model fails to describe thermoelectric systems other than standard ones, such as, e.g., a p-n inline configuration module; Downey and coworkers thus make use of a thermal transmission line model accounting explicitly for the contacts to overcome this limitation. Both models are equivalent for commercial modules \cite{Downey2007}. The possibility of distinguishing both thermal contacts in the thermal transmission line model offers similar features as the general expression given in Eq.~(\ref{Zth}). Hence, with this equation, it is possible to recover the surprising diminution of $|{\mathcal Z}|$ below its high frequencies limit, i.e., $R$, for the single TE leg of a p-type sample made of Na$_{0.8}$Pb$_{20}$Sb$_{0.6}$Te$_{22}$ studied in \cite{Downey2007}.

More recently, De Marchi and Giaretto have also recovered a similar RC behavior for the thermal contribution to the electrical small signal impedance using a lumped parameters model \cite{DeMarchi2011}. Yet, contrary to Downey and coworkers, they linked the characteristic angular frequency, designated as $\omega_{p\ell}$ in Ref.~\cite{DeMarchi2011}, to the thermal properties of the system. They found that $\omega_{p\ell} = K_0 / C_{\rm C}$. This expression should be compared to the characteristic angular frequency $\omega_{\rm TE}$ given in Eq.~(\ref{omega}): De Marchi and Giaretto neglected in the derivation of $\omega_{p\ell}$ both the thermal conductance of the contacts and the thermal capacitance of the thermoelectric element. These simplifications compared to Eq.~(\ref{omega}) might however be justified as, on the one hand, adiabatic conditions at the edges of the system are assumed in Ref.~\cite{DeMarchi2011}, leading to $K_{\rm C}=0$, and, on the other hand, the thermal capacitance of the active part of the module $C_{\rm th}$ is about one order of magnitude lower than that of the thermal contacts $C_{{\rm C}}$ for commercial thermoelectric modules (see, e.g., the example considered in Ref.~\cite{Lineykin2005}). Equation~(\ref{omega}) thus is an extension of the result of De Marchi and Giaretto allowing relaxation of the adiabatic conditions during measurements used in previous studies, e.g., Refs.~\cite{DeMarchi2011,Garcia2014}. Without such constrain, it is then possible to envisage impedance spectroscopy as an appropriate technique to perform \emph{in-situ} characterization of thermoelectric modules, including determination of the thermal contact quality. Note that a method based on the dc contribution of $R_{\rm TE}$ has already been proposed to evaluate thermal contact conductances~\cite{Sim2015}. 

Consideration of both capacitances $C_{\rm th}$ and $C_{\rm C}$ in the expression of the characteristic angular frequency also sheds light in recent results by Garc\'ia-Ca\~nadas and Min. In Ref.~\cite{Garcia2014}, the characteristic angular frequency is given by $\omega_{\rm TE} = K_0 / C_{\rm th}$. The value for the thermal capacitance of the thermoelectric module extracted using that expression is about one order of magnitude larger than typical values for the material used in the module (Bi$_2$Te$_3$). This discrepancy is easily overcome if one uses Eq.~(\ref{omega}): The thermal capacitance obtained by Garc\'ia-Ca\~nadas and Min is not $C_{\rm th}$ alone but $C_{\rm th}+C_{\rm C}$. As already stressed, for commercial thermoelectric module $C_{\rm C}$ is about one order of magnitude larger than $C_{{\rm th}}$, so the high value obtained in Ref.~\cite{Garcia2014} is likely due to the contribution of $C_{\rm C}$. Figure~\ref{fig:figure4}, where the frequency response of a system with neglected contact thermal capacitances is shown, illustrates that this assumption leads to overestimate the characteristic angular frequency $\omega_{\rm TE}$ while the impedance magnitude variation associated with $R_{\rm TE}$ is left unchanged. This point is thus consistent with the fact that the thermal capacitance $C_{\rm C}$ is the only parameter with a significant discrepancy.

While the simple expression of the electrical impedance ${\mathcal Z}_{\rm sym}$ obtained for the symmetric case, i.e., Eq.~(\ref{Zsimple}), describes properly the main feature of the small signal electrical response of thermoelectric modules, it fails to explain the slight modification of the response at frequencies higher than $\omega_{\rm TE}$ observed experimentally in Refs.~\cite{DeMarchi2011,Garcia2014,Garcia2014JEM}. This small discrepancy compared to the simple RC model is particularly visible on a Nyquist plot where the half circle associated with Eq.~(\ref{Zsimple}) is deformed, leading De Marchi and Giaretto to coin the term ``porcupine diagram'' to designate this specific diagram~\cite{DeMarchi2011}. In their analysis, these authors consider the different thermal responses of the active thermoelectric part, of the electrical insulator, typically a ceramic, but also of the electrical conductors used to connect the different legs of the module together. In our model, the thermal contacts encompass these two latter layers. Garc\'ia-Ca\~nadas and Min managed however to recover the ``snoot of the porcupine'' without distinguishing the two layers in the thermal contacts \cite{Garcia2014}. While it is possible to reproduce such a diagram using Eq.~(\ref{Zth}), the associated values for the contrast functions seem quite unrealistic for commercial modules, i.e., $\Psi_{\rm K}\approx1$ and $\Psi_{\rm C}\approx0$. The asymmetry of the contacts thus seems not to be responsible for the modification of the higher frequencies response. As pointed out in Ref.~\cite{DeMarchi2011}, this anomaly cannot be evidenced with a global approach such as the one used in the present article: One needs to start from the local heat equation.

\subsection{On the frequency dependence of the thermoelectric properties}

We end this discussion with the frequency limitation of the model developed in the present article. The value of the different parameters $R$, $K_0$ and $\alpha$ obtained in the dynamical regime might differ from those in the continuous regime \cite{Ezzahri2014}, these latter being the ones needed to evaluate performances. It is thus important to consider in practice sufficiently low frequencies to limit this discrepancy. For example, noticeable variations of the thermoelectric properties of $\rm Si_{0.7}Ge_{0.3}$ appear above 1~MHz \cite{Ezzahri2014}; these are mainly due to the electrical properties of this material. It appears however that characteristic frequencies $\omega_{\rm TE}/(2\pi)$ due to temperature fluctuations remain below 1~Hz \cite{Downey2007,DeMarchi2011,Garcia2014JEM}. The assumption of constant parameters regarding frequency thus is meaningful for analysis of impedance spectroscopy measurements.

\section{Conclusion}

In this article, we introduced a small-signal model for thermoelectric generator. While recovering the form of the equivalent electrical impedance of the system empirically given by Downey et coworkers \cite{Downey2007}, we derived this model on a physical basis, extending an existing dc model to the dynamical regime. This model is particularly appealing as it allows to consider non vanishing contact thermal conductance between the thermoelectric system and its environment, which is very important for optimization purposes amd performance improvement \cite{Elghool2017,Araiz2017}; further, this constraint relaxation compared to previous models paves the way to in-situ characterization of such devices through impedance spectroscopy measurements. This model might also be extended to account for more complicated device combination configurations, electrically and thermally in series and/or parallel where interface effects can have a non-negligible impact on the device operation \cite{Apertet2012brief,Apertet2015}.

\section*{Acknowledgments}
H. Ouerdane is supported by the Skoltech NGP Program (Skoltech-MIT joint project). We are pleased to acknowledge Prof.~C.~Goupil for the suggestion of the problem treated in this paper and for careful reading of the manuscript. We are also pleased to thank Dr. J. Garc\'{\i}a-Ca\~nadas for valuable comments and suggestions. 

\section*{References}

\bibliographystyle{elsarticle-num}

\end{document}